# InN and GaN/InN monolayers grown on ZnO {0001}

Torsten Ernst[1,2], Caroline Chèze[1,*], Raffaella Calarco[1]

[1]*Paul-Drude-Institut für Festkörperelektronik, Hausvogteiplatz 5-7, 10117 Berlin, Germany*

[2]*TopGaN Laser, Sokołowska 29-37, 01-142 Warsaw, Poland*

Thin InN and GaN/InN films were grown on oxygen-polar (O) $(000\bar{1})$ and zinc-polar (Zn) (0001) zinc oxide (ZnO) by plasma-assisted molecular beam epitaxy (PAMBE). The influence of the growth rate (GR) and the substrate polarity on the growth mode and the surface morphology of InN and GaN/InN was investigated *in situ* by reflection *high-energy* electron diffraction (RHEED) and *ex situ* by atomic force microscopy (AFM). During InN deposition, a transition from two dimensional to three dimensional (2D-3D) growth mode is observed in RHEED. The critical thickness for relaxation increases with decreasing GR and varies from 0.6 ML (GR: 1.0 ML/s) to 1.2 MLs (GR: 0.2 ML/s) on O-ZnO and from 1.2 MLs (GR: 0.5 ML/s) to 1.7 MLs (GR: 0.2 ML/s) on Zn-ZnO. The critical thickness for relaxation of GaN on top of 1.2 MLs and 1.5 MLs thick InN is close to zero on O-ZnO and 1.6 MLs on Zn-ZnO, respectively.

*Electronic mail: cheze@pdi-berlin.de

## I. Introduction

The band gap of (In,Ga)N alloys covers a wide spectral range from ultraviolet (GaN) to infrared (InN). These alloys are widely used as the active region of optoelectronic devices like light-emitting diodes or laser diodes.[1] Yet, the growth of (In,Ga)N alloys with high indium (In) content as required to access longer emission wavelengths is still challenging. The large difference in atomic radius of In and gallium (Ga) causes several strain-related issues. One of these is the composition pulling effect - which describes a compositional gradient along the growth direction.[2] Another strain-related issue is the phase separation of (In,Ga)N due to spinodal decomposition,[3] leading to nano-scale inhomogeneities not only in the In-content but also in the thickness of (In,Ga)N multiple quantum well (MQW) structures. These inhomogeneities cause the localization of charge carriers at regions of higher In-content or wider well thickness, which translates into a broadening of the emission spectrum and energy-dependent radiative decay times.[4,5]

As a way to circumvent the immiscibility of (In,Ga)N, short period superlattices (SPSLs) made of alternating binary InN and GaN layers, so-called digital alloys, were proposed.[6] Nonetheless, the fabrication of such digital alloys is difficult because it involves the combination of InN and GaN that have a large difference in optimum growth temperature,[7,8,9] and lattice constant.[10] Following the results of Yoshikawa *et al.*[6] and Dimakis *et al.*[11] who reported the successful fabrication of such digital alloys on Ga-polar GaN, several other groups uncovered a discrepancy between the intended InN/GaN SPSLs and the (In,Ga)N/GaN SPSLs with an In-content of 33% at maximum that were obtained instead.[12,13,14] This discrepancy is consistent with theoretical calculations[15] predicting the absence of any growth window for coherent InN on GaN. Recently, Lymperakis *et al.* concluded that strain causes the elastic frustrated rehybridisation of In atoms on the surface in a $(2\sqrt{3}\times2\sqrt{3})R30°$ structure. Therefore, the In-incorporation in (In,Ga)N may be limited by strain induced by the substrate. However, Duff *et al.* also indicated the emergence of a narrow growth window for metal-polar InN on a more lattice matched substrate such as metal-polar $In_{0.25}Ga_{0.75}N$ under growth conditions of high In and N excess, or on N-polar GaN.[15] Therefore, to overcome the limitation in In content in (In,Ga)N, a possible strategy is to choose a substrate that offers a smaller lattice mismatch with InN than the value of 11% for the growth on GaN.

ZnO is an attractive substrate material because it is isomorphic to (In,Ga)N and lattice-matched to $In_{0.18}Ga_{0.82}N$.[16] Moreover, III-N may be grown on ZnO with desired polarity, the Zn-ZnO surface yielding the metal-polarity of III-N, while the O-ZnO might yield mostly metal polar.[17] This finding enables already the fabrication of InN films of high structural quality synthesized on O-face $ZnO(000\bar{1})$[18] despite the still large lattice mismatch of 8.9%.[10,19] However, any introduction of Ga in InN even in amount as small as 5% induces high defect densities and a severe degradation of the interface between (In,Ga)N and $ZnO(000\bar{1})$, due to a high chemical reactivity of Ga with ZnO.[20] Therefore in this study, we investigate the growth of InN and GaN with a thickness of 6 MLs on top of a thin InN layer on ZnO of both the Zn- and the O-polarities. We developed a protocol to protect the ZnO surface from Ga by depositing a ML-thick coherent InN buffer layer on ZnO prior to the GaN growth.

**II. Experimental Section**

The growth was conducted in a DCA P600 plasma-assisted molecular beam epitaxy (PAMBE) system equipped with standard effusion cells and a radio frequency plasma source to supply active nitrogen.

We utilized commercial *c*-plane ZnO substrates from CrysTec polished on both faces: Zn-ZnO (0001) and O-ZnO (000$\bar{1}$). The ZnO substrates were annealed in a resistance heat oven at 1050°C.[21] Smooth O-ZnO surfaces [Figure 1 (a)] were obtained in most cases after annealing of the substrate at 1050°C for 1 hour in $O_2$ atmosphere of 0.5 bar and at a flux of 150 sccm. Nevertheless, some three dimensional (3D) protrusions occasionally occurred. Those clusters were removed by rinsing in de-ionized water for 20 minutes under sonication and renewed annealing of the samples. For Zn-ZnO surfaces annealing at 1050°C for 1 hour with 1 bar of $O_2$ atmosphere and a flux of 400 sccm resulted in smooth, stepped surfaces [Figure 1 (b)], with a root mean square roughness (RMS) measured by atomic force microscopy (AFM) on 5×5 μm$^2$ areas of 0.25 nm or less.

For all samples the growth temperature was set to 450°C as measured by a thermocouple. Nitrogen-rich growth conditions during InN deposition (N/In~1.1) and stoichiometric growth conditions N/Ga~1 during GaN growth were chosen to prevent the excess of metal at the surface which may etch the ZnO substrate.[20] In-fluxes of 0.2 ML/s, 0.5 ML/s, and 1.0 ML/s were used, while the Ga flux was set to 0.2 ML/s. The layers were deposited on both ZnO faces in separate experiments. A first set of samples consists of InN layers of 1 ML, 2 MLs, and 10 MLs thickness deposited on the smooth ZnO surface of the annealed substrates. A second set of samples consists of 6.0 MLs of GaN deposited subsequently on top of both 1.2 MLs of InN on O-ZnO (000$\bar{1}$) and 1.5 MLs of InN on Zn-ZnO (0001).

We monitored the growth *in situ* by reflection *high-energy* electron diffraction (RHEED) and we investigated the morphological properties of the samples *ex situ* by AFM.

Before all growth experiments, the investigation of the bare ZnO substrates by RHEED revealed a streaky pattern on both the Zn-ZnO and the O-ZnO polarities, which corresponds well to the smooth surface that we have prepared following our annealing procedures. Importantly, *in situ* measurement by quadrupole mass spectrometry in the line of sight revealed that metal desorption was negligible at the low growth temperature employed. In addition, no metal droplets were found on the surface after growth and the surface revealed by AFM was clear. Thus, we infer that all supplied In and Ga amounts were incorporated into the layers.

## III. Results

### A. Influence of growth rate and polarity on the growth of ML thick InN

For all of these experiments, RHEED monitoring revealed a transition from a streaky pattern at the start of the InN deposition to a spotty one in the course of InN deposition larger than 1 ML. [Fig. 2(a), (b), (c), (d)] exemplarily illustrates these patterns before and after the deposition of InN. Next, we have monitored the evolution of the *a*-lattice spacing by measuring the streak spacing during the deposition of InN. The results are shown in Fig. 2(e) and (f) for the growth on O-ZnO $(000\bar{1})$ and on Zn-ZnO (0001), respectively. For all growth rates (GRs), as growth starts, the lattice spacing shows only a slight increase for the samples grown on O-ZnO while it remains unchanged for growth on Zn-ZnO. This observation indicates that InN grows at first nearly pseudomorphic to ZnO for both polarities. However, beyond a critical layer thickness the lattice spacing increases, i.e. relaxation sets in. On both polarities, the critical layer thickness increases with decreasing GR and is fairly comparable (cf. Table I). The largest critical thickness measured for the lowest growth rate is 1.2 MLs on O-ZnO and 1.7 MLs on Zn-ZnO.

**TABLE I.** Critical layer thickness for relaxation ($h_{crit}$) and for the transition from 2D to 3D growth mode ($h_{2D-3D}$) of InN thin films on ZnO.

| GR (ML/s) | $h_{crit}$ O-ZnO | $h_{crit}$ Zn-ZnO | $h_{2D-3D}$ O-ZnO | $h_{2D-3D}$ Zn-ZnO |
|---|---|---|---|---|
| 1.0 | 0.6 ML |  | 1.2 MLs |  |
| 0.5 | 1.0 ML | 1.2 MLs | 1.0 ML | 1.2 MLs |
| 0.2 | 1.2 MLs | 1.7 MLs | 1.2 MLs | 1.7 MLs |

The roughening of the surface revealed by the emergence of the spotty patterns in RHEED occurs almost at the beginning of the relaxation process for all samples, except for the sample grown on O-ZnO at the highest GR. In this case the 2D to 3D transition occurs at 1.2 MLs while the relaxation starts at 0.6 ML [see Table I and Fig. 2(e) and (f)].

The observation of the surface roughening is confirmed by the evolution of the specular reflection intensity in RHEED as exemplarily shown in Fig. 2(g) for the growth of InN on O-ZnO with GR 1.0 ML/s. After a slight intensity decrease, the signal increases for the whole growth. This increase in intensity coincides with the 2D to 3D growth mode transition. The same behavior is observed for the growth of InN on Zn-ZnO (Fig. 2(h)). This observation is also confirmed by the surface morphology of four dedicated samples investigated by AFM [Fig. 1 (c), (d), (e), (f)]. These four samples resulted from the deposition of only 1.0 ML and 2.0 MLs of InN on O-ZnO and 1.0 ML and 10.0 MLs of InN on Zn-ZnO at a GR of 0.2 ML/s.

After the deposition of 1.0 ML of InN on both O-ZnO and Zn-ZnO, smooth surfaces covered by mono-atomic steps and a roughness of about 0.3 – 0.4 nm are obtained [Fig. 1(c) and (d)]. However, after the deposition of 2.0 MLs of InN on O-ZnO [Fig. 1(e)], the surface roughness increases to 1.6 nm. Though mono-atomic steps are still observed, additional 3D structures formed on top, with a density of $2.2 \times 10^8$ cm$^{-2}$, and height and diameter of 17 nm and 64 nm, respectively. For the Zn-ZnO polarity, after the deposition of 10 MLs 3D structures were revealed with a density of $5.7 \times 10^6$ cm$^{-2}$ and height and diameter of 5 nm and 190 nm, respectively. Thus these results clarify the streaky-spotty pattern transition observed in RHEED, and most importantly, they reveal that we can grow coherent and smooth InN layers with thickness of about 1.7 MLs (1.2 MLs) on Zn-ZnO (O-ZnO). In the following we utilize smooth thin InN layers as an intermediate protective layer for the growth of (In,Ga)N/GaN heterostructures in order to prevent chemical reactions between Ga and ZnO. InN thickness was chosen to be below both h$_{crit}$ and h$_{2D-3D}$.

## B. Deposition of 6.0 MLs of GaN on coherently grown InN layers

We next focus on the growth of GaN on InN/ZnO. In agreement with the above results, we have obtained a streaky RHEED pattern after the deposition at GR 0.2 ML/s of InN layers with thickness 1.2 MLs on O-ZnO and 1.5 MLs on Zn-ZnO. However, the streaky pattern changes to a spotty one in the course of GaN deposition on both ZnO polarities [Fig. 3(a), (b)]. For growth on O-ZnO the RHEED streaks decrease in intensity (not shown) until they almost disappear and simultaneously faint spots appear [Fig. 3(a)]. This effect occurs at a total thickness of 1.3 MLs, i.e. it corresponds to the beginning of the GaN deposition. For growth on Zn-ZnO, the same intensity evolution is observed but the 2D – 3D transition is delayed to a total thickness of 4.6 MLs, i.e. it occurs after the deposition of 3.1 MLs GaN.

The variation of the *a*-lattice spacing in dependence of the InN and GaN thicknesses deposited on O-ZnO and Zn-ZnO is presented in Figures 3(c) and (d), respectively. Depending on the ZnO polarity, a very different evolution occurs.

On O-ZnO [Fig. 3(c)], the lattice spacing increases during the deposition of 1.2 MLs of InN and gradually decreases immediately at the start of GaN deposition. Beyond the deposition of 1.3

MLs of GaN the lattice spacing even decreases further below the lattice spacing of ZnO and it stabilizes at the value of 3.24 Å after the total deposition of about 5 MLs (1.2 MLs of InN + 3.8 MLs of GaN). This value corresponds to a remaining lattice mismatch of 1.3% for GaN. In order to understand this phenomenon, we have considered the possible alloying of the supplied InN and GaN as an (In,Ga)N layer. We considered two cases: unstrained and strained (In,Ga)N. Therefore, we also display in Figure 3(c) the evolution of the lattice spacing of an unstrained (In,Ga)N layer that was calculated following Vegard's law. In that first case, the calculated lattice spacing clearly does not follow the measured one. Nonetheless, in the second case where the InN *a*-lattice spacing is strained to the experimental value attained just before the start of GaN deposition (equal to 3.31 Å), we obtain a satisfactory agreement between the experimental and calculated values up to a total layer thickness of 4.2 MLs (1.2 MLs InN + 3 MLs GaN). Beyond 4.2 MLs, the experimental lattice spacing decreases slower than the model. Hence, alloying does not seem to be the mechanism at play to explain the experimental data. Therefore, we examined a third case fitting the experimental data by an exponential decay law [dotted curve in Fig. 3(c)] following a model introduced by Bourret *et. al.* [22] to describe the evolution of the lattice strain in GaN/AlN heterostructures, measured by RHEED. This model is used in this work to describe the relaxation behavior of GaN/InN/ZnO heterostructures. This law reads:

$$a(h) = a_1 exp\left(\frac{-(h-h_c)}{\Lambda}\right) + a_0 \qquad (1)$$

where *a(h)* is the *a*-lattice spacing as a function of the layer thickness $h$, $a_1$ is the total amplitude of the lattice relaxation, $h_c$ is a critical thickness, $\Lambda$ is a characteristic thickness at which a fraction 1-1/e of the plastic relaxation has occurred, and $a_0$ is the unstrained reference lattice constant of the film. This model provides a nice fit of the experimental values for the fitting parameters $a_1$=0.07 Å, $a_0$=3.23 Å, $h_c$=0.2 ML and $\Lambda$=0.9 ML.

In striking contrast, on Zn-ZnO [Fig. 3(d)] the *a*-lattice spacing remains constant up to a total layer thickness of around 3.1 MLs (1.5 MLs InN + 1.6 MLs GaN). Thus, the thickness of 1.6 MLs can be assigned to the critical thickness of GaN on InN fully strained to Zn-ZnO. Beyond this thickness the *a*-lattice spacing increases, which indicates the formation of partially relaxed (In,Ga)N. However, after 4.6 MLs (1.5 MLs InN + 3.1 MLs GaN) the *a*-lattice spacing inverts its evolution and decreases. The streaky-spotty transition of the RHEED pattern occurs at around 4.6 MLs starting simultaneously to the onset of the *a*-lattice decrease.

As expected from the spotty RHEED patterns that emerged during both experiments, AFM measurements of these GaN/InN/ZnO heterostructures revealed the presence of 3D features on the sample surface [Fig. 1(g) and (h)]. For growth on O-ZnO the density, average height, and diameter of these 3D features are $7.1 \times 10^8$ cm$^{-2}$, 27.5 nm, and 43.5 nm, respectively. The RMS roughness value for this sample is 4.4 nm. For growth on Zn-ZnO the density, average height, and diameter of the 3D features are $5.6 \times 10^8$ cm$^{-2}$, 10.3 nm, and 55.4 nm, respectively. The RMS roughness value for this sample is 3.2 nm. The total volume of these 3D features is for both samples similar with 1.4 MLs (Zn-ZnO) and 1.5 MLs (O-ZnO). However, in both cases the surface between them is still very smooth as atomic steps are observed [Fig. 1(g) and (h)].

In short, we have established that for growth on Zn-ZnO, templates consisting of 1.5 MLs InN and 1.6 MLs GaN can be pseudomorphically grown. For growth on O-ZnO no pseudomorphic growth of InN/GaN templates could be established. In this case sub-ML InN already shows relaxation with an increasing *a*-lattice constant, and for GaN grown on 1.2 MLs InN the *a*-lattice constant decreases right after the growth starts with an immediate streaky-spotty transition of the RHEED pattern.

## IV. Discussion

For the deposition of InN layers, our results evidenced that the relaxation seems to follow the Stranski-Krastanov mechanism with critical thickness that depends not only on the ZnO polarity but also on the GR. This last effect has not been reported so far for III-N. For InAs on InP, Nakayama *et. al.* reported the increase of the critical thickness with increasing GR.[23] He attributed this finding to the slow transformation of a metastable 2D film, growing faster as the growth rate is increased, into stable 3D islands. . In contrast, in the present case of InN on ZnO, we observe a slight delay of relaxation for lower GR. Following the interpretation of Nakayama, this result would rather indicate the faster rate of transformation from the 2D film into 3D islands with increasing GR. As shown above, whatever the ZnO polarity, an almost pseudomorphic 2D growth sets in at the start of the InN deposition. In RHEED, the intensity of the specular reflection decreases at the onset of InN deposition on both ZnO polar surfaces, which may result from the nucleation of platelets with height 1 – 2 MLs similarly to the growth of GaN/AlN heterostructures.[22] This process develops until a specific $h_{crit}$ that depends on the GR used for the InN deposition. Once $h_{crit}$ is attained, the strain relaxation process initiates with an enlargement of the in-plane lattice parameter, as monitored by RHEED. However, it then proceeds differently for each ZnO polarity.

For growth of InN on both polarities the relaxation in almost all cases occurs close to the transition from 2D to 3D growth mode. Relaxation might occur first elastically and later on plastically with introduction of dislocations,[24] or synchronously with dislocations. For InN on GaN, islands usually form after strain has been already partially relieved by dislocations. However, Figure 1(e) and (f) evidence the presence of small pits of hexagonal shape surrounding the nanoscale islands. Such morphology was reported already in InN layers grown on O-ZnO and originates from the disturbed InN/ZnO interface where the ZnO substrate is severely etched down.[18] The islands were shown to have a crystal structure that matches the one of cubic $In_2O_3$.[18] Thus for layer thickness exceeding $h_{crit}$ in the order of a ML, defects form in the InN adlayer and as relaxation occurs, the InN/ZnO interface is no longer protected towards back etching that leads to the formation of $In_2O_3$ inclusions similarly to the step erosion mechanism occurring during InAs quantum dot formation on GaAs.[25]

For the case of growth of InN on O-ZnO with GR 1 ML/s, in contrast, the relaxation generates within the 2D growth mode. Therefore, the coalescence of the platelets should occur with the concomitant formation of dislocations as suggested by the simultaneous increase of the lattice parameter in RHEED for InN thickness above 0.6 ML [Fig. 2(e)]. Such relaxation mechanisms might be due to the formation of dislocations at the junction of 2D platelets,[22] however further investigations to elucidate the details of the process are necessary.

The critical thickness for the 2D – 3D growth mode transition, $h_{2D-3D}$, is governed by a balance between the surface energy and the elastic energy.[24] As a consequence, depending on growth conditions, surface polarity, and / or growth techniques $h_{2D-3D}$ might vary. The observation that both $h_{crit}$ and $h_{2D-3D}$ depend on MBE growth parameters suggests a kinetic origin of the film growth mode with possible strong influences from film morphology.[26]

For the deposition of GaN on InN pseudmorphic to ZnO, the relaxation process is also characteristic of the polarity of ZnO. On InN/O-ZnO the GaN lattice spacing immediately decreases and occurs simultaneously to the 2D-3D transition observed in RHEED. Therefore, at the low temperature of 450°C GaN growth seems to develop according to the Volmer-Weber mode on the 1 ML-thick InN layer. Nonetheless, the overgrowth of the unstable strained InN layer by the GaN cap layer could also possibly have triggered the sharp 2D-3D transition with increasing coverage.[27] Considering the formation and the relaxation of GaN 3D islands we obtain a good fit to the *a*-lattice spacing evolution. However, the formation of strained or unstrained (In,Ga)N might also occur during capping by GaN,[28] although the presence of (In,Ga)N alone does not allow to properly fit the data.

Interestingly, the relaxation of the 6.0 ML-thick GaN layer is not full and the GaN *a*-lattice spacing reaches a steady-state *a*-lattice spacing around 3.24 Å that corresponds either to GaN with a residual strain of 1.3% or to fully relaxed $In_{0.12}Ga_{0.88}N$.

In contrast on In-polar InN/Zn-ZnO the growth of GaN is pseudomorphic until a total layer thickness of about 3.1 MLs (1.5 MLs InN + 1.6 MLs GaN) is reached. Afterwards the *a*-lattice constant increases. Such a behavior is unexpected and might be due to composition pulling effect that allows the formation of (In,Ga)N with a higher In-content on the surface. In fact, RHEED describes transient of surface lattice spacing during growth. The relatively high strain state of the pseudomorphic layer allows for the composition pulling effect to appear. At a thickness of 4.6 MLs the *a*-lattice spacing starts to decrease and concomitantly the 2D-3D transition occurs similarly to InN/ZnO.

**V. Conclusions**

Our investigations showed that the critical layer thickness for relaxation of InN grown on ZnO is dependent on the GR and the ZnO-polarity. InN grown on O-ZnO exhibited a lower critical layer thickness than InN grown on Zn-ZnO. Within the range of our investigations – with GR between 0.2 ML/s and 1.0 ML/s – the critical layer thickness decreased with increasing GR. We determined the critical layer thickness to be between 0.6 ML (GR: 1.0 ML/s) and 1.2 MLs (GR: 0.2 ML/s) on O-ZnO and 1.2 MLs (GR: 0.5 ML/s) and 1.7 MLs (GR: 0.2 ML/s) on Zn-ZnO. This considerably limits the thickness of an InN buffer layer that is supposed to prevent chemical reactions between subsequently deposited Ga atoms and ZnO. For almost all samples grown the critical layer thickness upon which a 2D-3D growth regime transition takes place is similar to the critical layer thickness of relaxation. This sets an upper limit for the thickness of InN to be used in MQW or SPSL structures grown on ZnO.

The effects of the deposition of GaN with a GR of 0.2 ML/s on 1.2 ML (O-ZnO) or 1.5 ML (Zn-ZnO) thin InN buffer layers was investigated. For growth on InN/O-ZnO the results show that shortly after commencing the deposition of GaN, 3D structures formed. For growth on InN/Zn-ZnO 3D structures start to form for GaN deposited thickness beyond 3.1 MLs. This suggests that the growth of InN/GaN MQWs and InN/GaN SPSLs may not be possible on O-ZnO under the growth conditions investigated in this study, while for growth on Zn-ZnO, the formation of (1 ML (In,Ga)N)/(1-2 MLs (In,Ga)N) SPSLs structures might be possible, with different In-contents in the well and barrier respectively.

We thank R.B. Lewis for a critical reading of the manuscript and H. P. Schönherr for MBE maintenance. Funding of this work by the European Union's Horizon 2020 research and innovation program (Marie Skłodowska-Curie Actions) under grant agreement "SPRInG" No. 642574 is gratefully acknowledged.

Appl. Phys. **50**, 031004 (2011).

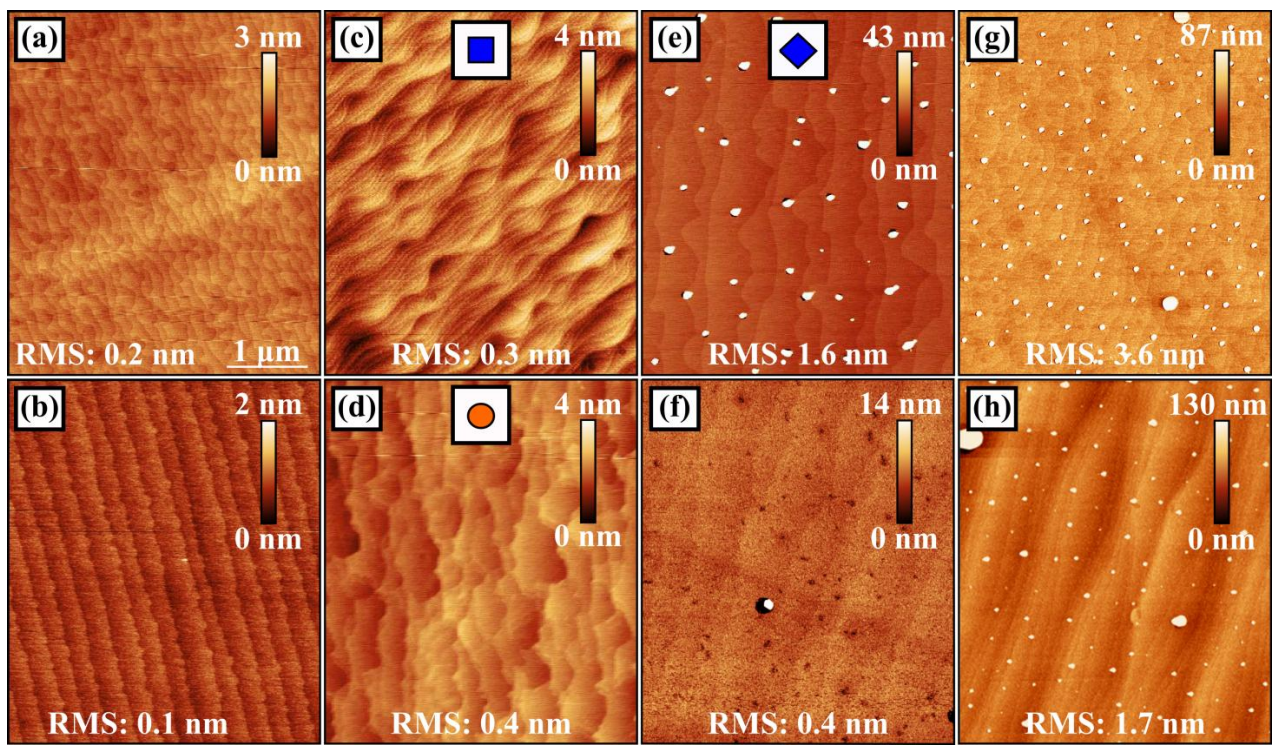

**Fig. 1.** AFM images of size 5 μm x 5 μm: (a) O-ZnO and (b) Zn-ZnO substrates after annealing, (c) InN(1.0 ML)/O-ZnO and (d) InN(1.0 ML)/Zn-ZnO, (e) InN(2.0 MLs)/O-ZnO and (f) InN(10.0 MLs)/Zn-ZnO, (g) GaN(6.0 MLs)/InN(1.2 MLs)/O-ZnO and (h) GaN(6.0 MLs)/InN(1.5 MLs)/Zn-ZnO. Symbols in (c), (d) and (e) recall Figure 2 (e) and (f).

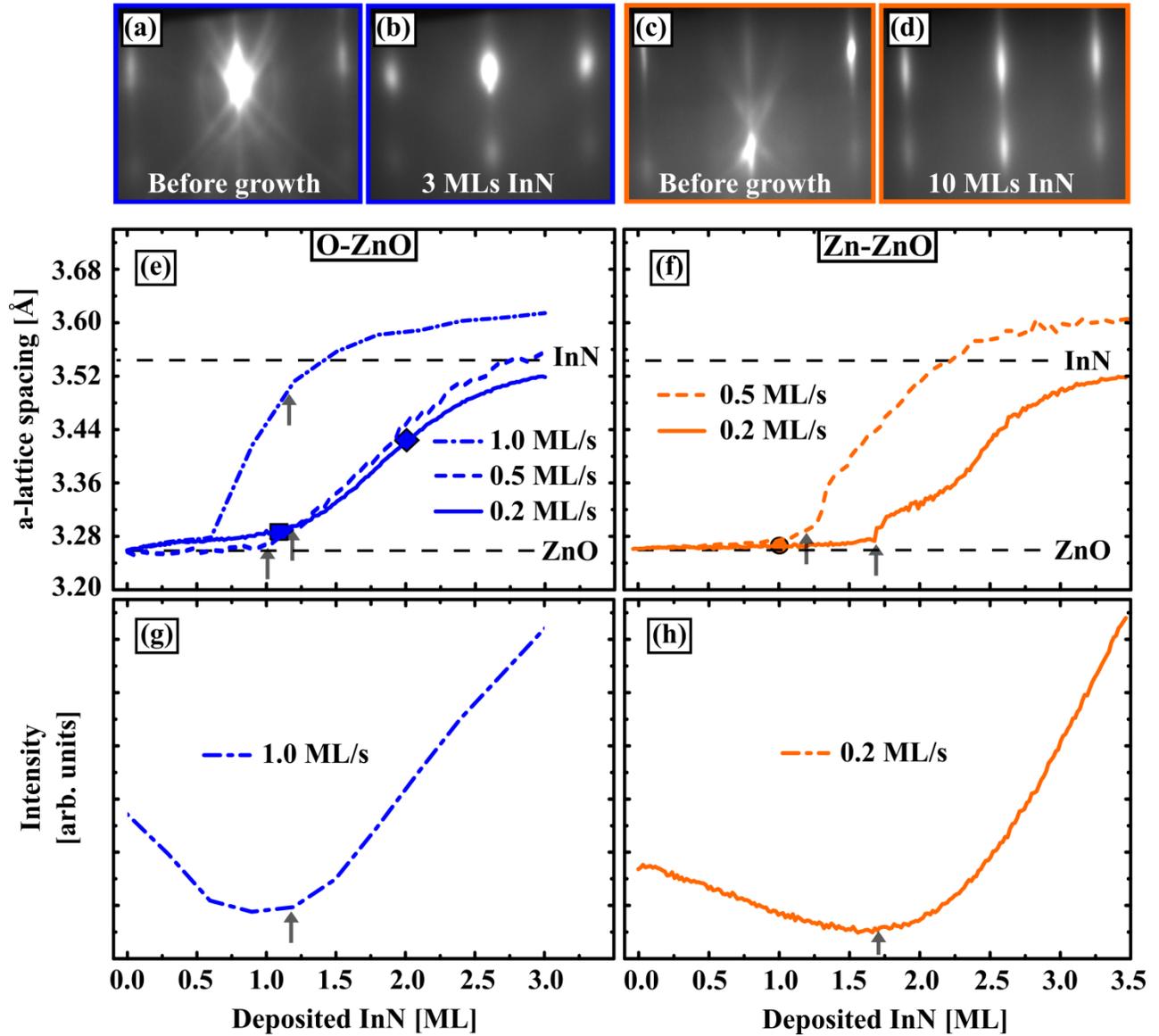

**Fig. 2**. RHEED patterns of (a), (b) O-ZnO before and after InN growth and (c), (d) Zn-ZnO before and after InN growth; (e) and (f) evolution of *a*-lattice spacing during the deposition of InN at different GR on O-ZnO and Zn-ZnO, respectively; (g), (h) intensity evolution of the RHEED specular spot of InN/O-ZnO with GR 1.0 ML/s and InN/Zn-ZnO with GR 0.2 ML/s. The grey arrows highlight the transition from streaky to spotty RHEED patterns corresponding to $h_{2D-3D}$, while the dashed black lines indicate the unstrained lattice constants of InN and ZnO. Symbols in (e) and (f) recall Figure 1(c), (d) and (e).

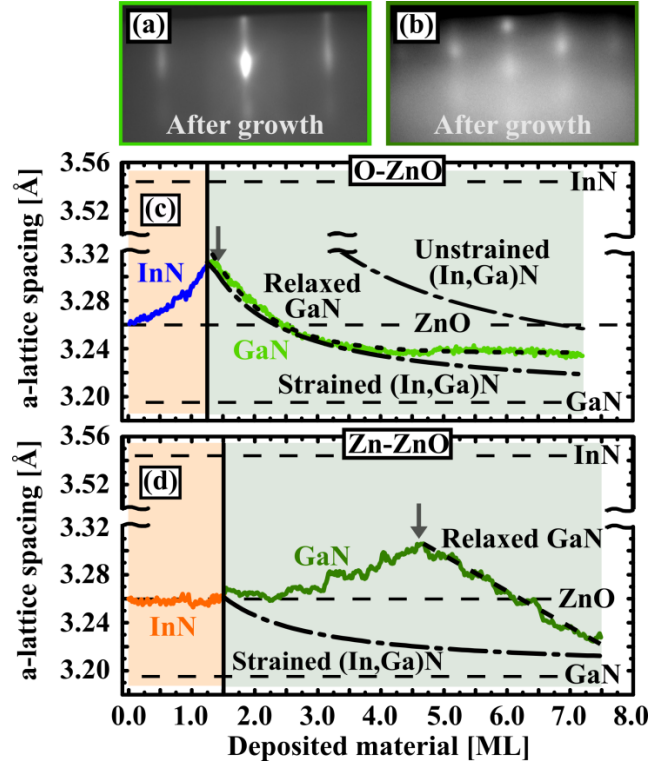

**Fig. 3**. (a) and (b) RHEED patterns in the $\langle 11\bar{2}0 \rangle$ and in the $\langle 10\bar{1}0 \rangle$ azimuths after the deposition of 6.0 MLs GaN on 1.2 MLs InN/O-ZnO and 1.5 MLs InN/Zn-ZnO, respectively. *a*-lattice spacing evolution during deposition of 6 MLs GaN on (c) 1.2 MLs InN on O-ZnO and (d) 1.5 MLs InN on Zn-ZnO. The grey arrows indicate the transition between streaky and spotty RHEED diffraction patterns ($h_{2D-3D}$), while the dashed black lines indicate the unstrained lattice constants of InN, GaN and ZnO. The black dashed/dotted curves in (c) and (d) correspond to (In,Ga)N layers following Vegard's law. The black dotted curve corresponds to partially relaxed GaN.